\documentclass[conference, 10pt]{IEEEtran}

\usepackage[cmex10]{amsmath}  
\usepackage{amssymb}
\usepackage{graphicx,color}   

\usepackage{ifxetex}
\ifxetex
\usepackage{xunicode}
\else
\usepackage[utf8]{inputenc}
\usepackage[T1]{fontenc}
\usepackage{microtype}
\fi

\newtheorem{theorem}{Theorem}
\newtheorem{remark}{Remark}

\newtheorem{corollary}{Corollary}

\begin{document}

\title{The Capacity Region of Restricted Multi-Way Relay Channels with Deterministic Uplinks}
\author{\IEEEauthorblockN{Lawrence Ong and Sarah J.\ Johnson}
\IEEEauthorblockA{School of Electrical Engineering and Computer Science, 
The University of Newcastle, Australia\\
Email: lawrence.ong@cantab.net, sarah.johnson@newcastle.edu.au }
}
\maketitle

\begin{abstract}
This paper considers the multi-way relay channel (MWRC) where multiple users exchange messages via a single relay. The capacity region is derived for a special class of MWRCs where (i) the uplink and the downlink are separated in the sense that there is no direct user-to-user links, (ii) the channel is restricted in the sense that each user's transmitted channel symbols can depend on only its own message, but not on its received channel symbols, and (iii) the uplink is any deterministic function.
\end{abstract}

\section{Introduction}

Two-way communications, where two nodes exchange messages, were first studied by Shannon~\cite{shannon61}. Though this channel is seemingly simple---with only two nodes---the capacity region remains unknown to date, except for a few spacial cases: the Gaussian two-way channel~\cite{han84} and the {\em restricted}\footnote{In the restricted channel, each user's transmitted symbols can depend only on its own message but not on its received symbols.} two-way channel~\cite{shannon61}.

One variation of the two-way channel is the two-way relay channel~\cite{rankovwittneben06} where the exchange of messages between two users is assisted by a relay, which itself has no message to send. The two-way relay channel has many applications including satellite communications and cellular mobile communications. However, the introduction of a relay in the channel further complicates the task of finding its capacity. The difficulty lies in determining the optimal processing at the relay. To focus on the relay, recent work considers the {\em separated} two-way relay channel where there is no direct user-to-user links~\cite{gunduztuncel08,namchunglee09}. This model is also motivated by real-life scenarios where relays are used when inter-user communications are not possible. However, the capacity region of even the restricted and separated two-way relay channel is not known in general~\cite{ongjohnson12cl}.

A natural extension of the two-way relay channel is the multi-way relay channel (MWRC) where multiple users exchange messages through a relay~\cite{gunduzyener09}. Apart from theoretical interests, this extension is also motivated by satellite networks and cellular networks with multiple users. To the best of our knowledge, the only class of the MWRCs where the capacity has been found is the separated finite-field MWRC~\cite{ongmjohnsonit11}. For the separated Gaussian MWRC, the capacity has been found for only the symmetrical case~\cite{ong10amwrc}.

In this paper, the capacity region is found for another class of MWRCs, where (i) the uplink and the downlink are separated, (ii) the channel is restricted, and (iii) the uplink is any {\em deterministic} function (there is no noise, but signals from different users interfere with each other). 
For this class of MWRCs, we will show that an optimal scheme is for the relay to directly map its received vector to a codeword to be transmitted. Each user then performs a two-step decoding: it first decodes the received vectors of the relay, and then decodes the messages of all other users.

\section{Main Results}

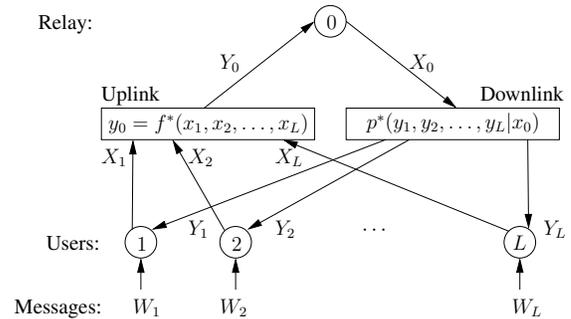
\begin{figure}[t]
\centering
\resizebox{7.5cm}{!}{ 
\begin{picture}(0,0)%
\includegraphics{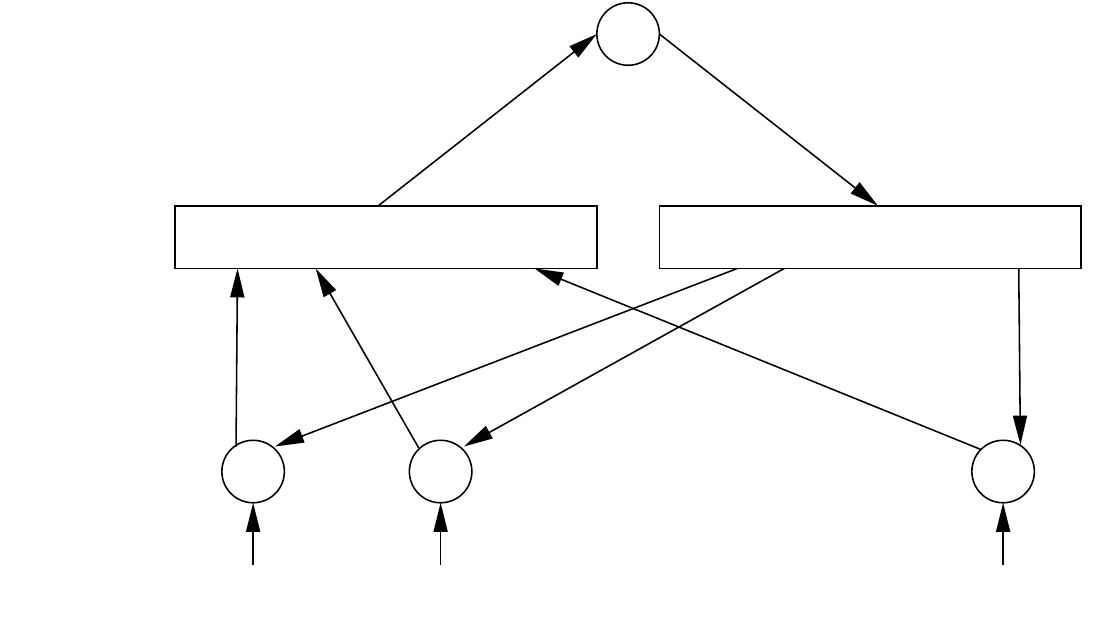}%
\end{picture}%
\setlength{\unitlength}{3947sp}%
\begingroup\makeatletter\ifx\SetFigFont\undefined%
\gdef\SetFigFont#1#2#3#4#5{%
 \fontsize{#1}{#2pt}%
  \fontfamily{#3}\fontseries{#4}\fontshape{#5}%
  \selectfont}%
\fi\endgroup%
\begin{picture}(5301,3010)(-614,-2764)
\put(2746,-931){\makebox(0,0)[lb]{\smash{{\SetFigFont{12}{14.4}{\familydefault}{\mddefault}{\updefault}{\color[rgb]{0,0,0}$p^*(y_1,y_2,\dotsc,y_L|x_0)$}%
}}}}
\put(564,-2093){\makebox(0,0)[lb]{\smash{{\SetFigFont{12}{14.4}{\familydefault}{\mddefault}{\updefault}{\color[rgb]{0,0,0}$1$}%
}}}}
\put(1463,-2093){\makebox(0,0)[lb]{\smash{{\SetFigFont{12}{14.4}{\familydefault}{\mddefault}{\updefault}{\color[rgb]{0,0,0}$2$}%
}}}}
\put(4134,-2085){\makebox(0,0)[lb]{\smash{{\SetFigFont{12}{14.4}{\familydefault}{\mddefault}{\updefault}{\color[rgb]{0,0,0}$L$}%
}}}}
\put(2364, 14){\makebox(0,0)[lb]{\smash{{\SetFigFont{12}{14.4}{\familydefault}{\mddefault}{\updefault}{\color[rgb]{0,0,0}$0$}%
}}}}
\put(3151,-361){\makebox(0,0)[lb]{\smash{{\SetFigFont{12}{14.4}{\familydefault}{\mddefault}{\updefault}{\color[rgb]{0,0,0}$X_0$}%
}}}}
\put(1351,-361){\makebox(0,0)[lb]{\smash{{\SetFigFont{12}{14.4}{\familydefault}{\mddefault}{\updefault}{\color[rgb]{0,0,0}$Y_0$}%
}}}}
\put(1876,-1261){\makebox(0,0)[lb]{\smash{{\SetFigFont{12}{14.4}{\familydefault}{\mddefault}{\updefault}{\color[rgb]{0,0,0}$X_L$}%
}}}}
\put(2701,-1936){\makebox(0,0)[lb]{\smash{{\SetFigFont{12}{14.4}{\familydefault}{\mddefault}{\updefault}{\color[rgb]{0,0,0}$\dotsm$}%
}}}}
\put(526,-2686){\makebox(0,0)[lb]{\smash{{\SetFigFont{12}{14.4}{\familydefault}{\mddefault}{\updefault}{\color[rgb]{0,0,0}$W_1$}%
}}}}
\put(1351,-2686){\makebox(0,0)[lb]{\smash{{\SetFigFont{12}{14.4}{\familydefault}{\mddefault}{\updefault}{\color[rgb]{0,0,0}$W_2$}%
}}}}
\put(4126,-2686){\makebox(0,0)[lb]{\smash{{\SetFigFont{12}{14.4}{\familydefault}{\mddefault}{\updefault}{\color[rgb]{0,0,0}$W_L$}%
}}}}
\put(226,-1261){\makebox(0,0)[lb]{\smash{{\SetFigFont{12}{14.4}{\familydefault}{\mddefault}{\updefault}{\color[rgb]{0,0,0}$X_1$}%
}}}}
\put(1051,-1261){\makebox(0,0)[lb]{\smash{{\SetFigFont{12}{14.4}{\familydefault}{\mddefault}{\updefault}{\color[rgb]{0,0,0}$X_2$}%
}}}}
\put(1051,-1936){\makebox(0,0)[lb]{\smash{{\SetFigFont{12}{14.4}{\familydefault}{\mddefault}{\updefault}{\color[rgb]{0,0,0}$Y_1$}%
}}}}
\put(1876,-1936){\makebox(0,0)[lb]{\smash{{\SetFigFont{12}{14.4}{\familydefault}{\mddefault}{\updefault}{\color[rgb]{0,0,0}$Y_2$}%
}}}}
\put(4426,-1936){\makebox(0,0)[lb]{\smash{{\SetFigFont{12}{14.4}{\familydefault}{\mddefault}{\updefault}{\color[rgb]{0,0,0}$Y_L$}%
}}}}
\put(-374, 14){\makebox(0,0)[lb]{\smash{{\SetFigFont{12}{14.4}{\familydefault}{\mddefault}{\updefault}{\color[rgb]{0,0,0}Relay:}%
}}}}
\put(-299,-2086){\makebox(0,0)[lb]{\smash{{\SetFigFont{12}{14.4}{\familydefault}{\mddefault}{\updefault}{\color[rgb]{0,0,0}Users:}%
}}}}
\put(-599,-2686){\makebox(0,0)[lb]{\smash{{\SetFigFont{12}{14.4}{\familydefault}{\mddefault}{\updefault}{\color[rgb]{0,0,0}Messages:}%
}}}}
\put(226,-661){\makebox(0,0)[lb]{\smash{{\SetFigFont{12}{14.4}{\familydefault}{\mddefault}{\updefault}{\color[rgb]{0,0,0}Uplink}%
}}}}
\put(3826,-661){\makebox(0,0)[lb]{\smash{{\SetFigFont{12}{14.4}{\familydefault}{\mddefault}{\updefault}{\color[rgb]{0,0,0}Downlink}%
}}}}
\put(293,-939){\makebox(0,0)[lb]{\smash{{\SetFigFont{12}{14.4}{\familydefault}{\mddefault}{\updefault}{\color[rgb]{0,0,0}$y_0=f^*(x_1,x_2,\dotsc,x_L)$}%
}}}}
\end{picture}%
}
\caption{The MWRC with a deterministic uplink and an arbitrary downlink, where $f^*(\cdot)$ is a deterministic function and $p^*(\cdot)$ is a probability distribution function}
\label{fig:mwrc}
\end{figure}

\subsection{Notation}

Random variables are denoted by upper-case letters. For a random variable $X$, the lower-case letter $x$ denotes its realization, and the script letter $\mathcal{X}$ denotes its alphabet. Subscripts are used to denote the node to which the symbol belongs and the time index of the symbol, e.g., $X_{it}$ is the channel input from node $i$ at time $t$.

For collections of symbols, we use bold face to denote the sequence of a symbol from time $t=1$ to $t=n$, e.g., $\boldsymbol{X}_i \triangleq (X_{i1}, X_{i2}, \dotsc, X_{in})$, and subscripts in brackets to denote symbols from a group of nodes, e.g., let $\mathcal{A} = \{1,2,4\}$, then $X_{(\mathcal{A})} = (X_1, X_2,X_4)$. The set of integers from 1 to $N$ inclusive is denoted by $[1:N] \triangleq \{1,2,\dotsc, N\}$. So, for example, $X_{([1:L])} = (X_1,X_2,\dotsc, X_L)$.

\subsection{Channel Model}

The MWRC consists of $L$ users (denoted by node $1, 2, \dotsc, L$) and one relay (denoted by node $0$). We denote by $x_i$ the input to the channel from node $i$, and by $y_i$ the channel output received by node $i$, for all $i \in [0:L]$. We also denote by $w_j$ user $j$'s message, for all $j \in [1:L]$. The MWRC is depicted in Fig.~\ref{fig:mwrc}.

The {\em separated} MWRC is denoted by $2(L+1)$ finite sets $\mathcal{X}_0, \mathcal{X}_1, \dotsc, \mathcal{X}_L, \mathcal{Y}_0, \mathcal{Y}_1, \dotsc, \mathcal{Y}_L$, and two probability distribution functions (pdf): (i) the uplink $p^*(y_0|x_1,x_2,\dotsc,x_L)$, and (ii) the downlink $p^*(y_1,y_2,\dotsc,y_L|x_0)$. 
The channel is separated in the sense that there is no direct user-to-user link. 

In this paper, we consider MWRCs with {\em deterministic uplink} where the pdf for the uplink has the following form:
\begin{equation*}
p^*(y_0|x_1,x_2,\dotsc,x_L) = \begin{cases}
1, & \text{if } y_0 = f^*(x_1,x_2,\dotsc,x_L) \\
0, & \text{otherwise}.
\end{cases} 
\end{equation*}
Here, $f^*(\cdot)$ is a deterministic function. Note that this deterministic model includes the linear finite-field deterministic channels~\cite{avestimehrsezgin10}  as a special case. Also note that we do not impose any restriction on the downlink.

A $(2^{nR_1}, 2^{nR_2}, \dotsc, 2^{nR_L}, n)$ code for the MWRC consists of the following: (i) One message for each user: $w_i \in \mathcal{W}_i = [1: 2^{nR_i}]$, for all $i \in [1:L]$. (ii) An encoding function for each user: $\boldsymbol{x}_i(w_i)$, $i \in [1:L]$. (iii) A set of encoding functions for the relay: $x_{0t} = g_t (y_{01}, y_{02}, \dotsc, y_{0t-1})$, for all $t \in [1:n]$. (iv) A decoding function at each user: $(\hat{w}_{1<i>}, \hat{w}_{2<i>}, \dotsc, \hat{w}_{i-1<i>}, \hat{w}_{i+1<i>}, \dotsc, \hat{w}_{L<i>}) = h_i(w_i,\boldsymbol{y}_i)$, where $\hat{w}_{j<i>}$ is the estimate of the user $j$'s message by user $i$. Using this code, each user $i$ transmits at $R_i$ bits/channel use.


The above code structure imposes the following constraints: (i) The channel is {\em restricted}: each user's transmit symbols can depend only on its message but not on its received symbols. (ii) The operation at the relay is {\em causal}: the relay's transmit symbol can only depends on its previously received symbols. (iii) The users engage in {\em full data exchange}: each user is to send its message to the other $(L-1)$ users.

Assuming that each message $W_i$ is uniformly distributed in $\mathcal{W}_i$, the average probability of error of the code is defined as $P_\text{e} = 2^{-n\sum_{i=1}^M R_i} \sum_{w_{([1:L])}} \Pr \{ \hat{W}_{j<i>} \neq w_j \text{ for some } j \neq i | W_{([1:L])} = w_{([1:L])} \}$. The rate tuple $(R_1,R_2,\dotsc,R_L)$ is said to be {\em achievable} if the following is true: for any $\eta > 0$, there exists for sufficiently large $n$ a $(2^{nR_1}, 2^{nR_2}, \dotsc, 2^{nR_L}, n)$ code such that $P_\text{e} < \eta$. The {\em capacity region} is the closure of the set of achievable tuples.


\subsection{Main Results}

Denote the set of non-negative real vectors of length-$L$ by $\mathbb{R}_+^L$, and define the following two sets of rate tuples:
$\mathcal{R}^* \triangleq \Bigg\{ (R_1,R_2\dotsc, R_L) \in \mathbb{R}_+^L:$\\
\begin{align}
&\sum_{j \in [1:L]\setminus \{i\}} R_j \leq I(X_0;Y_i),\quad \forall i \in [1:L], \label{eq:main-1}\\
&\sum_{j \in \mathcal{U}} R_j \leq H(Y_0 | X_{([1:L] \setminus \mathcal{U})}, Q), \quad \forall \text{ non-empty } \mathcal{U} \subset [1:L],\label{eq:main-2}
\end{align}
and for some joint distribution
\begin{multline}
p(q,x_{([0:L])}, y_{([0:L])})\\ = p(q) \prod_{i=1}^L p(x_i|q) p(x_0) p^*(y_0|x_{([1:L])}) p^*(y_{([1:L])}|x_0), \label{eq:input}
\end{multline}
where the cardinality of $Q$ is bounded as $|\mathcal{Q}| \leq L+1 \Bigg\},$\\
and $\mathcal{R}^\text{o}$ is $\mathcal{R}^*$ with the inequalities in \eqref{eq:main-1} and \eqref{eq:main-2} replaced by strict inequalities. Here, $Q$ is an auxiliary random variable.
We have the following results:

\begin{theorem} \label{theorem:outer-bound}
If a rate tuple $(R_1,R_2,\dotsc,R_L)$ is achievable, then $(R_1,R_2,\dotsc,R_L) \in \mathcal{R}^*$.
\end{theorem}

\begin{theorem} \label{theorem:inner-bound}
If a rate tuple $(R_1,R_2,\dotsc,R_L) \in \mathcal{R}^\text{o}$, then $(R_1,R_2,\dotsc,R_L)$ is achievable.
\end{theorem}

We will prove Theorem~\ref{theorem:outer-bound} in Section~\ref{section:outer} and Theorem~\ref{theorem:inner-bound} in Section~\ref{section:inner}.

Since the set of pdfs of the form \eqref{eq:input} is compact, and both $H(\cdot)$ and $I(\cdot)$ are continuous functions of the pdf, it follows that $\mathcal{R}^*$ is closed. So, the closure of $\mathcal{R}^\text{o}$ gives $\mathcal{R}^*$.
Combining Theorems~\ref{theorem:outer-bound} and \ref{theorem:inner-bound}, we have the following capacity result:
 \begin{theorem} \label{theorem:main}
The capacity region of the restricted and separated MWRC with a deterministic uplink is $\mathcal{R}^*$.
\end{theorem}

\subsection{A Special Case}


A special case of the MWRC with a deterministic uplink is where
\begin{align}
&|\mathcal{X}_j| \geq |\mathcal{X}_0|, \quad \text{for all }j \in [1:L] \label{eq:special-1}\\
&Y_0 = (X_1,X_2,\dotsc,X_L). \label{eq:special-2}
\end{align}
Condition \eqref{eq:special-1} implies that any $R_j$ that satisfies \eqref{eq:main-1} must also satisfy $R_j \leq I(X_0;Y_i) \leq H(X_0) \leq \log_2 |\mathcal{X}_0| \leq \log_2 |\mathcal{X}_j|$, for all $i \in [1:L]$. Condition \eqref{eq:special-2} implies that $H(Y_0|X_{([1:L] \setminus \mathcal{U})},Q) = H(X_{(\mathcal{U})}|Q)$. By choosing $|\mathcal{Q}|=1$ and each $X_i$ ($i \in [1:L]$) to be independent and uniformly distributed, we have that $H(Y_0|X_{([1:L] \setminus \mathcal{U})},Q) = H(X_{(\mathcal{U})}) = \sum_{j \in \mathcal{U}} \log_2 |\mathcal{X}_j|$. So, $\sum_{j \in \mathcal{U}} R_j \leq H(Y_0|X_{([1:L] \setminus \mathcal{U})},Q)$, i.e., any rate tuple that satisfies \eqref{eq:main-1} also satisfies \eqref{eq:main-2}.
For this special channel, Theorem~\ref{theorem:main} reduces to the following:

\begin{corollary} \label{corollary:main}
Consider a restricted MWRC satisfying \eqref{eq:special-1} and \eqref{eq:special-2}. The capacity region is
\begin{align}
\mathcal{R}' \triangleq \Big\{ &(R_1,R_2\dotsc, R_L) \in \mathbb{R}_+^L: \nonumber \\
& \sum_{j \in [1:L]\setminus \{i\}} R_j \leq I(X_0;Y_i),\quad \forall i \in [1:L], \\
& \text{for some } p(x_0, y_{([1:L])}) = p(x_0) p^*(y_{([1:L])}|x_0). \Big\} \nonumber
\end{align}
\end{corollary}
The rate region $\mathcal{R}'$ is a function of only the downlink variables. 

\begin{remark}
Condition~\eqref{eq:special-2} implies that the uplink is deterministic and its input signals are {\em non-interfering}, i.e., the relay receives all transmitted codewords $\boldsymbol{x}_{([1:L])}$ without error.  Due to condition \eqref{eq:special-1}, for any rate tuple in $\mathcal{R}'$, we have $|\mathcal{W}_i| = 2^{nR_i}\leq|\mathcal{X}_i|^n$ for all $i \in [1:L]$. This means for any rate tuple in $\mathcal{R}'$, we can choose an uplink code $\{\boldsymbol{x}_i(w_i)\}$ for each user $i$  such that the relay is able to decode all $w_i$'s without error. With this observation, we note that Corollary~\ref{corollary:main} is consistent with the result from~\cite{oechteringschnurr08} where (i) there are two users $L=2$, and (ii) the relay is given both user's message ($w_1,w_2$) a priori.
\end{remark}


\section{Capacity Outer Bound} \label{section:outer}

In this section, we prove Theorem~\ref{theorem:outer-bound} using cut-set arguments for multiterminal networks~\cite[pp.~589--591]{coverthomas06}. Let the set $\mathcal{S}$ be a subset of the nodes, $\mathcal{S} \subset [0:L]$, and let $\mathcal{S}^\text{c} \triangleq [0:L] \setminus \mathcal{S}$, where the sets $\mathcal{S}$ and $\mathcal{S}^\text{c}$ each contain at least one user, i.e., $\mathcal{S} \cap [1:L] \neq \emptyset$ and $\mathcal{S}^\text{c} \cap [1:L] \neq \emptyset$. We canonically define $R_0 = 0$. 

Using the fact that
\begin{itemize}
\item each message $W_i$ is uniformly distributed in $\mathcal{W}_i$, 
\item the messages $\{ W_i: i \in [1:L] \}$ are independent, 
\item each user wants to decode the messages of all other users, meaning that $P_{\text{error},n} \triangleq \Pr \{$a (more capable) user wrongly decodes $W_{(\mathcal{S})}$ from $\boldsymbol{Y}_{(\mathcal{S}^\text{c})}$ and $W_{(\mathcal{S}^\text{c})} \}$ must be made arbitrarily small (the user is {\em more capable} in the sense that it has access to $\boldsymbol{Y}_{(\mathcal{S}^\text{c})}$ and $W_{(\mathcal{S}^\text{c})}$ instead of just $\boldsymbol{Y}_i$ and $W_i$),
\item $X_{it}$ is a function of $W_i$ for each $i \in [1:L]$,
\item $X_{0t}$ is a function of $(Y_{01}, \dotsc, Y_{0t-1})$, and
\item the channel is memoryless, i.e., $ (W_{([1:L])}, X_{([0:L])1},$ $X_{([0:L])2}, \dotsc, X_{([0:L])t-1}, Y_{([0:L])1}, Y_{([0:L])2}, \dotsc, Y_{([0:L])t-1})$ $\rightarrow X_{([0:L])t} \rightarrow Y_{([0:L])t} $ forms a Markov chain,
\end{itemize}
we obtain the following: if $(R_1,R_2,\dotsc,R_L)$ is achievable, then we must have that
\begin{subequations}
\begin{align}
&n \sum_{j \in \mathcal{S}} R_j \leq \sum_{t=1}^n I( X_{(\mathcal{S})t} ; Y_{(\mathcal{S}^\text{c})t} | X_{(\mathcal{S}^\text{c})t}) + n\epsilon_n \label{eq:cut-set-first}\\
&= n\sum_{t=1}^n \frac{1}{n} I( X_{(\mathcal{S})t} ; Y_{(\mathcal{S}^\text{c})t} | X_{(\mathcal{S}^\text{c})t}, Q=t) + n\epsilon_n  \\
&=  n I( X_{(\mathcal{S})Q} ; Y_{(\mathcal{S}^\text{c})Q} | X_{(\mathcal{S}^\text{c})Q}, Q) + n\epsilon_n \\
&= n I( X_{(\mathcal{S})} ; Y_{(\mathcal{S}^\text{c})} | X_{(\mathcal{S}^\text{c})}, Q) + n\epsilon_n, \label{eq:change}
\end{align}
\end{subequations}
where \eqref{eq:cut-set-first} follows from ~\cite[eqns.~(15.321)--(15.332)]{coverthomas06} where $\epsilon_n \rightarrow 0$ as $P_{\text{error},n} \rightarrow 0$, $Q$ is a random variable which is uniformly distributed on $[1\,{:}\,n]$ and is independent of $(\boldsymbol{X}_{([1:L])},\boldsymbol{Y}_{([1:L])})$, and \eqref{eq:change} is derived by defining $\{X_i \triangleq X_{iQ}, \;Y_i \triangleq Y_{iQ}:\; i \in [1\, {:}\, L]\}$ whose distribution depends on $Q$ the same way as that of $(X_{{([1:L])}t},Y_{{([1:L])}t})$ on $t$. The RHS of \eqref{eq:change} is evaluated for some $p(q)p(x_{([0:L])}|q)p^*(y_0|x_{([1:L])}) p^*(y_{([1:L])}|x_0)$,  i.e., 
\begin{equation}
(Q, X_{([1:L])}) \rightarrow X_0 \rightarrow Y_{([1:L])} \label{eq:chain-2}
\end{equation}
forms a Markov chain. 

We now apply the cut-set bound~\eqref{eq:change} to the specific channel considered in this paper. Define $\mathcal{U} \triangleq \mathcal{S} \cap [1:L]$ and $\mathcal{U}^\text{c} \triangleq \mathcal{S}^\text{c} \cap [1:L] = [1:L] \setminus \mathcal{U}$. We consider two cases:

\noindent{\underline{Case 1: $0 \in \mathcal{S}^\text{c}$.}} For this case, we have $\mathcal{S} = \mathcal{U}$ and $\mathcal{S}^\text{c} = \{0\} \cup \mathcal{U}^\text{c}$. So,
\begin{subequations}
\begin{align}
&\sum_{j \in \mathcal{U}} R_j \leq I( X_{(\mathcal{U})} ; Y_0, Y_{(\mathcal{U}^\text{c})} |  X_0, X_{(\mathcal{U}^\text{c})}, Q)  + \epsilon_n\\
&= I(X_{(\mathcal{U})} ; Y_0 |  X_0, X_{(\mathcal{U}^\text{c})}, Q) \nonumber \\
&\quad + I(X_{(\mathcal{U})} ; Y_{(\mathcal{U}^\text{c})} |  X_0, X_{(\mathcal{U}^\text{c})}, Q, Y_0)  + \epsilon_n \label{eq:chain-rule} \\
&= I(X_{(\mathcal{U})} ; Y_0 |  X_0, X_{(\mathcal{U}^\text{c})}, Q)  + \epsilon_n \label{eq:markov-1} \\
&= H( Y_0 |  X_0, X_{(\mathcal{U}^\text{c})}, Q)  - H( Y_0 |  X_0, X_{(\mathcal{U}^\text{c})}, X_{(\mathcal{U})}, Q)  + \epsilon_n \label{eq:mutual} \\
&=  H( Y_0 |  X_0, X_{(\mathcal{U}^\text{c})}, Q)  + \epsilon_n \label{eq:deterministic} \\
&\leq H( Y_0 | X_{(\mathcal{U}^\text{c})}, Q)  + \epsilon_n, \label{eq:conditioning}
\end{align}
\end{subequations}
where \eqref{eq:chain-rule} follows from the chain rule, \eqref{eq:markov-1}  follows from the Markov chain \eqref{eq:chain-2}, \eqref{eq:mutual} follows from the definition of mutual information, \eqref{eq:deterministic} follows from the deterministic uplink, i.e., $Y_0$ is a deterministic function of $(X_{(\mathcal{U}^\text{c})}, X_{(\mathcal{U})})$, and \eqref{eq:conditioning} is obtained because conditioning cannot increase entropy.  Now, since $W_i$'s are independent, $\{ X_{it}(W_i): i \in [1:L]\}$ are independent for a fixed $t$.  It follows that $\{X_i: i \in [1:L]\}$ are independent given $Q$. So, \eqref{eq:conditioning} must hold for all selections of non-empty strict subsets $\mathcal{U} \subset [1:L]$ for some $p(q) \prod_{i=1}^L p(x_i|q) p^*(y_0|x_{(1:L)})$.

\noindent{\underline{Case 2: $0 \in \mathcal{S}$.}} For this case, we have $\mathcal{S} = \{0\} \cup \mathcal{U}$ and $\mathcal{S}^\text{c} = \mathcal{U}^\text{c}$. So,
\begin{subequations}
\begin{align}
\sum_{j \in \mathcal{U}} R_j  &\leq I( X_0, X_{(\mathcal{U})} ; Y_{(\mathcal{U}^\text{c})} | X_{(\mathcal{U}^\text{c})}, Q)  + \epsilon_n \\
&= H(Y_{(\mathcal{U}^\text{c})} | X_{(\mathcal{U}^\text{c})}, Q) - H(Y_{(\mathcal{U}^\text{c})} | X_{(\mathcal{U}^\text{c})}, Q, X_0, X_{(\mathcal{U})})\nonumber \\ &\quad + \epsilon_n \\
&= H(Y_{(\mathcal{U}^\text{c})} | X_{(\mathcal{U}^\text{c})}, Q) - H(Y_{(\mathcal{U}^\text{c})} |X_0) \nonumber \\ &\quad + I(Y_{(\mathcal{U}^\text{c})}; X_{[1:L]},Q |X_0) + \epsilon_n \\
&\leq H(Y_{(\mathcal{U}^\text{c})} ) - H(Y_{(\mathcal{U}^\text{c})} |X_0) + \epsilon_n \label{eq:markov-4}\\
&= I(X_0; Y_{(\mathcal{U}^\text{c})})  + \epsilon_n, \label{eq:markov-3}
\end{align}
\end{subequations}
where \eqref{eq:markov-4} is derived from the Markov chain \eqref{eq:chain-2} and because conditioning cannot increase entropy. 
Eqn.~\eqref{eq:markov-3} must hold for all selections of non-empty strict subsets $\mathcal{U} \subset [1:L]$ for some $p(x_0) p^*(y_{([1:L])} | x_0)$.

Now, we let $n \rightarrow \infty$ and $P_{\text{error},n} \rightarrow 0$ so that $\epsilon_n \rightarrow 0$.
Combining the results for Case 1 for all subsets $\mathcal{U}$ and Case 2 where $|\mathcal{U}^\text{c}|=1$, we have that if a rate tuple $(R_1,R_2,\dotsc,R_L)$ is achievable, then there exists some joint pdf of the form \eqref{eq:input} such that \eqref{eq:main-1} and \eqref{eq:main-2} hold.
This proves Theorem~\ref{theorem:outer-bound}. $\hfill\blacksquare$

\begin{remark}
Eqn.~\eqref{eq:change}, derived using cut-set arguments, is evaluated with some input distribution of the form $p(q)p(x_{([0:L])}|q)$. However, as we consider only constraints \eqref{eq:main-1} and \eqref{eq:main-2}, and as the messages $\{W_i\}$ are independent and the channel is restricted (i.e., users do not utilize feedback), it suffices to consider input distributions of the form $p(q) \prod_{i=1}^L p(x_i|q) p(x_0)$.
\end{remark}

\begin{remark}
The rate region lies in the $L$-dimensional Euclidean space. Hence, we can restrict the cardinality of the time-sharing random variable $Q$ to be $(L+1)$~\cite[p.~538]{coverthomas06}.
\end{remark}

\section{Capacity Inner Bound} \label{section:inner}

In this section, we prove Theorem~\ref{theorem:inner-bound}. The basic idea is that the relay creates a codebook that maps each possible unique sequence $\boldsymbol{y}_0$ that it receives to a unique codeword $\boldsymbol{x}_0$ to be transmitted. This technique is an extension to our previous work on two-way relay channels with deterministic uplinks~\cite{ongjohnson12cl}.

Consider $B$ blocks, each containing $n$ channel uses. Let each user transmit $(B-1)$ messages over these $B$ blocks, i.e., the messages of each user $i$ are $(W_i^{(1)}, W_i^{(2)}, \dotsc, W_i^{(B-1)})$ where each $W_i^{(b)} \in \mathcal{W}_i = [1:2^{nR_i}]$. If we can find coding schemes such that the probability that any user wrongly decode any messages can be made arbitrarily small, the rate tuple $\left( \frac{B-1}{B}R_1, \frac{B-1}{B}R_2, \dotsc, \frac{B-1}{B}R_L \right)$ is achievable.

 In the following sections, we consider the $b$-th block of uplink transmissions and the $(b+1)$-th block of downlink transmissions, in which each user $i$ sends $W_i^{(b)}$ and decodes $\{W_j^{(b)}$:  for all  $j \in [1:L] \setminus \{i\}\}$ and for some $b \in \{1,2,\dotsc, B-1\}$. For simplicity, we drop the subscript $b$.

\subsection{Codebook Generation}

Fix $p(q)$, $p(x_i|q)$ for all $i \in [1:L]$, and $p(x_0)$. Randomly generate a length-$n$ sequence $\boldsymbol{q}$ according to $\prod_{t=1}^n p(q_t)$. The vector $\boldsymbol{q}$ is made known to all users and the relay.

For each user $i \in [1:L]$, randomly and independently generate $2^{nR_i}$ length-$n$ sequences $\boldsymbol{x}_i$ according to $\prod_{t=1}^n p(x_{it}|q_t)$. Index the sequences $\boldsymbol{x}_i(w_i)$ for $w_i \in [1:2^{nR_i}]$. The codebook for user $i$ is $\mathcal{C}_i = \{ \boldsymbol{x}_i(w_i): w_i \in [1:2^{nR_i}]\}$.

Let $\overline{\mathcal{Y}_0^n}$ be the set of sequences $\boldsymbol{y}_0$ {\em induced} by the codewords of the users, i.e., $\overline{\mathcal{Y}_0^n} = \{ \boldsymbol{y}_0: \boldsymbol{x}_i \in \mathcal{C}_i, i \in [1:L] \}$. Since $\boldsymbol{y}_0$ is a deterministic function of $(\boldsymbol{x}_1, \boldsymbol{x}_2, \dotsc, \boldsymbol{x}_L$), the size of $\overline{\mathcal{Y}_0^n}$ is upper  bounded by $|\overline{\mathcal{Y}_0^n}| \leq \prod_{i=1}^L |\mathcal{C}_i|= 2^{n\sum_{i=1}^LR_i}$. Define $V = |\overline{\mathcal{Y}_0^n}|$.

For the relay, randomly and independently generate $V$ sequences $\boldsymbol{x}_0$ according to $\prod_{t=1}^n p(x_{0t})$. Index the sequences by $\boldsymbol{x}_0(v)$ for $v \in [1:V]$. The codebook of the relay is denoted by $\mathcal{C}_0 = \{ \boldsymbol{x}_0(v): v \in [1:V]\}$.

\subsection{Encoding}

In the $b$-th block, user $i$ transmits $\boldsymbol{x}_i(w_i) \in \mathcal{C}_i$, where $w_i$ is the message of user $i$. Let the received symbols at the relay be $\boldsymbol{y}_0 \in \overline{\mathcal{Y}_0^n}$. The relay defines a bijective mapping $\phi: \overline{\mathcal{Y}_0^n} \rightarrow [1:V]$. The relay transmits $\boldsymbol{x}_0(\phi(\boldsymbol{y}_0)) \in \mathcal{C}_0$ in the $(b+1)$-th block.

\subsection{Decoding for Each User}

Let the received symbols of user $i$ in block $(b+1)$ be $\boldsymbol{y}_i$. User $i$ attempts to decode the other users' messages sent in block $b$. 

Without loss of generality, assume that the transmitted messages are $w_i = a_i$. The relay receives the deterministic function $\boldsymbol{y}_0 = f^*( \boldsymbol{x}_1(a_1), \boldsymbol{x}_2(a_2), \dotsc, \boldsymbol{x}_L(a_L))$. Let $a_0 = \phi( \boldsymbol{y}_0 )$ be the corresponding index transmitted by the relay.

Define the following:
\begin{align*}
\mathcal{D}_{i}(a_i) \triangleq \Big\{ & v \in [1:V]: \\
&v = \phi( f^*( \boldsymbol{x}_1(w_1), \boldsymbol{x}_2(w_2), \dotsc, \boldsymbol{x}_L(w_L))), \\
& \text{where } w_i = a_i, \text{and } w_j \in [1:2^{nR_j}] \text{ for all } j \neq i. \Big\}
\end{align*}
This is the set of all possible indices sent by the relay given that the message of user $i$ is $a_i$. Clearly,
\begin{equation}
|\mathcal{D}_{i}(a_i)| \leq \prod_{j \in [1:L] \setminus \{i\}} 2^{nR_j} = 2^{n\sum_{j \in [1:L] \setminus \{i\}} R_j}. \label{eq:set-count}
\end{equation}

Each user decodes other user's messages in two steps:

\underline{Step 1: User $i$ decodes the index sent by the relay.} User $i$ declares that $\hat{v}_{<i>}$ is sent by the relay if $\hat{v}_{<i>}$ is the unique index such that $\hat{v}_{<i>} \in \mathcal{D}_i(a_i)$, and that $\boldsymbol{x}_0(\hat{v}_{<i>})$ and $\boldsymbol{y}_i$ are jointly typical, i.e., $(\boldsymbol{x}_0(\hat{v}_{<i>}), \boldsymbol{y}_i) \in \mathcal{A}_\epsilon^{(n)}(X_0,Y_i)$, where $\mathcal{A}_\epsilon^{(n)}(X_0,Y_i)$ is the set of jointly typical sequences~\cite[p.~195]{coverthomas06}. Otherwise, it declares an error. User $i$ makes an error in decoding the relay's index if any of the following events occurs:
\begin{itemize}
\item $\mathcal{E}_{1<i>}$: The correct index is not chosen, i.e., $a_0 \notin \mathcal{D}_i(a_i)$ or $(\boldsymbol{x}_0(a_0), \boldsymbol{Y}_i) \notin \mathcal{A}_\epsilon^{(n)}(X_0,Y_i) \}$.
\item $\mathcal{E}_{2<i>}$; Some wrong index is chosen, i.e., $(\boldsymbol{x}_0(a_0'), \boldsymbol{Y}_i) \in \mathcal{A}_\epsilon^{(n)}(X_0,Y_i)$, for some $a_0' \in \mathcal{D}_i(a_i) \setminus \{a_0\}$.
\end{itemize}

By definition, $a_0 \in \mathcal{D}_i(a_i)$. It follows from the joint asymptotic equipartition property (AEP)~\cite[p.~197]{coverthomas06} that $\Pr\{ \mathcal{E}_{1<i>} \}=\Pr \{ (\boldsymbol{x}_0(a_0), \boldsymbol{Y}_i) \notin \mathcal{A}_\epsilon^{(n)}(X_0,Y_i) \} \leq \epsilon$. Now,
\begin{subequations}
\begin{align}
&\Pr \{ \mathcal{E}_{2<i>} \} \nonumber \\
& \leq \sum_{ a_0' \in \mathcal{D}_i(a_i) \setminus \{ a_0\} } \Pr \{ (\boldsymbol{x}_0(a_0'), \boldsymbol{Y}_i) \in \mathcal{A}_\epsilon^{(n)}(X_0,Y_i) \} \label{eq:union} \\
& \leq \left(2^{n\sum_{j \in [1:L] \setminus \{i\}} R_j} -1 \right) 2^{-n( I(X_0;Y_i) - 3 \epsilon)} \label{eq:count-2} \\
& < 2^{n ( \sum_{j \in [1:L] \setminus \{i\}} R_j - I(X_0;Y_i) + 3\epsilon)}, \label{eq:error-event-2}
\end{align}
\end{subequations}
where \eqref{eq:union} follows from the union bound, \eqref{eq:count-2} follows from \eqref{eq:set-count} and the joint AEP~\cite[Thm.~7.6.1]{coverthomas06}. So, if
\begin{equation}
\sum_{j \in [1:L] \setminus \{i\}} R_j \leq I(X_0;Y_i) - 4\epsilon,  \label{eq:inner-bound-1}
\end{equation}
then $\Pr \{ \mathcal{E}_{2<i>} \} < 2^{-n\epsilon}$.

\underline{Step 2: User $i$ decodes the other users' messages.} Assume that user $i$ has decoded the relay's index $a_0$ (in Step 1) correctly. 
Knowing $a_0$, user $i$ obtains $\boldsymbol{y}_0 = \phi^{-1} (a_0)$. User $i$ declares that $\hat{w}_{j<i>}$ is sent by user $j$, for all $j \in [1:L] \setminus \{i\}$, if they are the unique messages such that
\begin{multline}\Big(\boldsymbol{q}, \boldsymbol{x}_1(\hat{w}_{1<i>}), \boldsymbol{x}_2(\hat{w}_{2<i>}), \dotsc, \boldsymbol{x}_{i-1}(\hat{w}_{i-1<i>}), \boldsymbol{x}_i(a_i), \\ \boldsymbol{x}_{i+1}(\hat{w}_{i+1<i>}), \boldsymbol{x}_{i+2}(\hat{w}_{i+2<i>}), \dotsc, \boldsymbol{x}_L(\hat{w}_{L<i>}), \boldsymbol{y}_0 \Big) \\ \in  \mathcal{A}_\epsilon^{(n)} (Q,X_1,X_2,\dotsc, X_L,Y_0). \label{eq:jointly-typical}
\end{multline}
Otherwise, it declares an error. User $i$ makes a decoding error if any of the following events occurs for some non-empty subset $\mathcal{U} \subseteq [1:L] \setminus \{i\}$:
\begin{itemize}
\item $\mathcal{E}_{0<i>}$: The correct sequences are not jointly typical, i.e., $\hat{w}_{j<i>} = a_j$ for all $j \in [1:L] \setminus \{i\}$ and \eqref{eq:jointly-typical} is not true.
\item $\mathcal{E}_{\mathcal{U}<i>}$: Some wrong sequences are jointly typical, i.e., \eqref{eq:jointly-typical} is true for (i) some $\hat{w}_{j<i>} \neq a_j$ for all $j \in \mathcal{U}$, and (ii) $\hat{w}_{k<i>} = a_k$ for all $k \in [1:L] \setminus ( \mathcal{U} \cup \{i\})$.
\end{itemize}
\begin{remark}
The error events here are similar to those for the multiple-access channel~\cite[pp.~4-27--4-28]{elgamalkim10notes}.
\end{remark}

Recall that $\mathcal{U}^\text{c} = [1:L] \setminus \mathcal{U}$. By joint AEP, we have that $\Pr \{ \mathcal{E}_{0<i>} \} \leq \epsilon$. In addition,
\begin{subequations}
\begin{align}
\Pr \{ \mathcal{E}_{\mathcal{U}<i>} \} &\leq \sum_{\substack{ \hat{w}_{j<i>} \in \mathcal{W}_j \setminus \{a_j\} \\ \forall j \in \mathcal{U} } }\Pr \{ \eqref{eq:jointly-typical} \text{ is true} \}\\
& \leq \prod_{j \in \mathcal{U}} (2^{nR_j}-1) 2^{-n(I(X_{(\mathcal{U})};Y_0,X_{(\mathcal{U}^\text{c})}|Q) - 6\epsilon)} \label{eq:joint-aep-2} \\
& = \prod_{j \in \mathcal{U}} (2^{nR_j}-1) 2^{-n( I(X_{(\mathcal{U})};Y_0|X_{(\mathcal{U}^\text{c})},Q)- 6\epsilon)} \label{eq:independent-2} \\
& < 2^{n ( \sum_{j \in \mathcal{U}} R_j - I(X_{(\mathcal{U})};Y_0|X_{(\mathcal{U}^\text{c})},Q) + 6\epsilon)},  
\end{align}
\end{subequations}
where \eqref{eq:joint-aep-2} follows from \cite[Thm.~15.2.3]{coverthomas06}, \eqref{eq:independent-2} is obtained because $X_{(\mathcal{U})}$ and $X_{(\mathcal{U}^\text{c})}$ are independent given $Q$. So, if
\begin{align}
\sum_{j \in \mathcal{U}} R_j &\leq  I(X_{(\mathcal{U})};Y_0|X_{(\mathcal{U}^\text{c})},Q) -  7\epsilon.  \nonumber\\
&= H(Y_0 | X_{(\mathcal{U}^\text{c})},Q) - H(Y_0| X_{(\mathcal{U})},X_{(\mathcal{U}^\text{c})},Q)   -  7\epsilon \nonumber \\
&= H(Y_0 | X_{(\mathcal{U}^\text{c})},Q)   -  7\epsilon, \label{eq:inner-bound-2}
\end{align}
then $\Pr \{ \mathcal{E}_{\mathcal{U}<i>} \}  < 2^{-n \epsilon}$.

\subsection{Decoding for All Users}

Now, we repeat the above decoding steps for all users $j \in [1:L]$, and for all blocks $b \in [1:B-1]$. Denote the probability that some user makes a decoding error in block $(b+1)$ by $P_{\text{e},b}$. From the union bound, we have
\begin{align}
P_{\text{e},b} &\leq \sum_{i \in [1:L]}  \Bigg[ \Pr \{ \mathcal{E}_{1<i>} \}  + \Pr \{ \mathcal{E}_{2<i>} \} + \Pr \{ \mathcal{E}_{0<i>} \} \nonumber \\
&\quad\quad\quad\quad\;\;  +\sum_{\substack{ \mathcal{U} \subseteq [1:L] \setminus \{i\} \\ \text{s.t. } |\mathcal{U}| \geq 1}}  \Pr \{ \mathcal{E}_{\mathcal{U}<i>} \} \Bigg], \\
P_\text{e} & \leq \sum_{b \in [1:B-1]} P_{\text{e},b} = (B-1) P_{\text{e},b}.
\end{align}

Suppose that a rate tuple $(R_1,R_2,\dotsc, R_L)$ satisfies \eqref{eq:main-1} and \eqref{eq:main-2} with strict inequalities for some joint pdf \eqref{eq:input}.
By choosing a sufficiently large $B$ and a sufficiently small $\epsilon$, we can satisfy $\sum_{j \in [1:L] \setminus \{i\}} \frac{B}{B-1}R_j + 4\epsilon \leq I(X_0;Y_i)$ each $i \in [1:L]$, and $\sum_{j \in \mathcal{U}} \frac{B}{B-1}R_j + 7 \epsilon \leq H(Y_0 | X_{(\mathcal{U}^\text{c})},Q)$ for each non-empty $\mathcal{U} \subset [1:L]$. By choosing a sufficiently large $n$, $P_\text{e} < (B-1) L \left( 2\epsilon + 2^{-n \epsilon} + (2^{L-1}-1) 2^{-n \epsilon} \right)$  can be made as small as desired. Hence the rate $(R_1,R_2,\dotsc,R_L)$ is achievable. This proves Theorem~\ref{theorem:inner-bound}. $\hfill \blacksquare$

\section{Further Remarks}

The achievable rate region in Theorem~\ref{theorem:inner-bound} has the same form as that of the noisy network coding (NNC) scheme~\cite{limkimelgamal11}. However, the NNC scheme allows the transmitted symbols of each user to depend on its previously received symbols---this is not allowed in the channel considered in this paper. Nevertheless, one can incorporate this restriction in the derivation of the NNC scheme and set $\hat{Y}_0=Y_0$, $\hat{Y}_i = \varnothing, \forall i \in [1:L]$ to obtain the same rate region in Theorem~\ref{theorem:inner-bound}. A major difference between the coding scheme in this paper and the modified NNC scheme is that, for the latter, the decoding of the messages in all $B$ blocks is done simultaneously after the entire $B$ blocks of transmission, and is therefore more complex and incurs a larger decoding delay.

A crucial point for the coding scheme derived in this paper to be optimal is that each user can recover $\boldsymbol{y}_0$ even when the codebook size of the relay is upper bounded by $2^{n\sum_{i \in [1:L]} R_i}$. This no longer holds when a small amount of noise is injected into the uplink, i.e., when $y_0$ is not a deterministic function of $x_{([1:L])}$. Hence this scheme might not be optimal for MWRCs with noisy uplinks.

Our coding scheme is derived for the restricted MWRC in which the users are not allowed to use feedback in their transmission. It has been shown~\cite{fongyeung11} that for a two-user MWRC with deterministic uplink and downlink, the capacity region of the unrestricted channel (where the transmitted symbols of the users can be functions of their previously received symbols) is strictly larger than the restricted channel. Hence, the coding scheme derived in this paper might not be optimal for the unrestricted MWRCs, even with deterministic uplinks.


\begin{thebibliography}{10}
\providecommand{\url}[1]{#1}
\csname url@samestyle\endcsname
\providecommand{\newblock}{\relax}
\providecommand{\bibinfo}[2]{#2}
\providecommand{\BIBentrySTDinterwordspacing}{\spaceskip=0pt\relax}
\providecommand{\BIBentryALTinterwordstretchfactor}{4}
\providecommand{\BIBentryALTinterwordspacing}{\spaceskip=\fontdimen2\font plus
\BIBentryALTinterwordstretchfactor\fontdimen3\font minus
  \fontdimen4\font\relax}
\providecommand{\BIBforeignlanguage}[2]{{%
\expandafter\ifx\csname l@#1\endcsname\relax
\typeout{** WARNING: IEEEtran.bst: No hyphenation pattern has been}%
\typeout{** loaded for the language `#1'. Using the pattern for}%
\typeout{** the default language instead.}%
\else
\language=\csname l@#1\endcsname
\fi
#2}}
\providecommand{\BIBdecl}{\relax}
\BIBdecl

\bibitem{shannon61}
C.~E. Shannon, ``Two-way communication channels,'' in \emph{Proc. 4th Berkeley
  Symp. Math. Stat. Probab.}, vol.~1.\hskip 1em plus 0.5em minus 0.4em\relax
  Univ. California Press, 1961, pp. 611--644.

\bibitem{han84}
T.~S. Han, ``A general coding scheme for the two-way channel,'' \emph{IEEE
  Trans. Inf. Theory}, vol. IT-30, no.~1, pp. 35--44, Jan. 1984.

\bibitem{rankovwittneben06}
B.~Rankov and A.~Wittneben, ``Achievable rate regions for the two-way relay
  channel,'' in \emph{Proc. IEEE Int. Symp. Inf. Theory (ISIT)}, Seattle, USA,
  July 9--14 2006, pp. 1668--1672.

\bibitem{gunduztuncel08}
D.~G{\"u}nd{\"u}z, E.~Tuncel, and J.~Nayak, ``Rate regions for the separated
  two-way relay channel,'' in \emph{Proc. 46th Allerton Conf. Commun. Control
  Comput. (Allerton Conf.)}, Monticello, USA, Sept. 23--26 2008, pp.
  1333--1340.

\bibitem{namchunglee09}
W.~Nam, S.~Chung, and Y.~H. Lee, ``Capacity of the {G}aussian two-way relay
  channel to within $\frac{1}{2}$ bit,'' \emph{IEEE Trans. Inf. Theory},
  vol.~56, no.~11, pp. 5488--5494, Nov. 2010.

\bibitem{ongjohnson12cl}
L.~Ong and S.~J. Johnson, ``The capacity region of the restricted two-way relay
  channel with any deterministic uplink,'' \emph{IEEE Commun. Lett.}, vol.~16,
  no.~3, pp. 396--399, Mar. 2011.

\bibitem{gunduzyener09}
D.~G{\"u}nd{\"u}z, A.~Yener, A.~Goldsmith, and H.~V. Poor, ``The multi-way
  relay channel,'' in \emph{Proc. IEEE Int. Symp. Inf. Theory (ISIT)}, Seoul,
  Korea, June 28--July 3 2009, pp. 339--343.

\bibitem{ongmjohnsonit11}
L.~Ong, S.~J. Johnson, and C.~M. Kellett, ``The capacity region of multiway
  relay channels over finite fields with full data exchange,'' \emph{IEEE
  Trans. Inf. Theory}, vol.~57, no.~5, pp. 3016--3031, May 2011.

\bibitem{ong10amwrc}
L.~Ong, C.~M. Kellett, and S.~J. Johnson, ``Capacity theorems for the {AWGN}
  multi-way relay channel,'' in \emph{Proc. IEEE Int. Symp. Inf. Theory
  (ISIT)}, Austin, USA, June 13--18 2010, pp. 664--668.

\bibitem{avestimehrsezgin10}
A.~S. Avestimehr, A.~Sezgin, and D.~N.~C. Tse, ``Capacity of the two-way relay
  channel within a constant gap,'' \emph{Eur. Trans. Telecomm.}, vol.~21,
  no.~4, pp. 363--374, June 2010.

\bibitem{oechteringschnurr08}
T.~J. Oechtering, C.~Schnurr, and H.~Boche, ``Broadcast capacity region of
  two-phase bidirectional relaying,'' \emph{IEEE Trans. Inf. Theory}, vol.~54,
  no.~1, pp. 454--458, Jan. 2008.

\bibitem{coverthomas06}
T.~M. Cover and J.~A. Thomas, \emph{Elements of Information Theory},
  2nd~ed.\hskip 1em plus 0.5em minus 0.4em\relax Wiley-Interscience, 2006.

\bibitem{elgamalkim10notes}
\BIBentryALTinterwordspacing
A.~{El Gamal} and Y.~Kim. (2010, June 22) Lecture notes on network information
  theory. [Online]. Available: \url{http://arxiv.org/abs/1001.3404v4}
\BIBentrySTDinterwordspacing

\bibitem{limkimelgamal11}
S.~H. Lim, Y.~Kim, A.~{El Gamal}, and S.~Chung, ``Noisy network coding,''
  \emph{IEEE Trans. Inf. Theory}, vol.~57, no.~5, pp. 3132--3152, May 2011.

\bibitem{fongyeung11}
S.~L. Fong and R.~W. Yeung, ``Feedback enlarges capacity region of two-way
  relay channel,'' in \emph{Proc. IEEE Int. Symp. on Inf. Theory (ISIT)}, St
  Petersburg, Russia, July 31--Aug. 5 2011, pp. 2248--2252.

\end{thebibliography}

\end{document}